\documentclass[aps,prb,preprint,groupedaddress]{revtex4-1}
\usepackage[dvipdfm]{}
\usepackage{graphicx}
\usepackage{epstopdf}
\usepackage{mathrsfs}
\usepackage{amsmath}
\def\tc{$T_{\rm c}$}
\def\tl{$1/T_1$} 
\def\kb{$k_{\rm B}T_{\rm c}$}

\def\pb{PbTaSe$_2$}

\begin{document}

\title{
Fully gapped spin-singlet superconductivity in noncentrosymmetric \pb\ : $^{207}$Pb-NMR
}
\author{S. Maeda$^{1}$, K. Matano$^1$, and Guo-qing Zheng$^{1,2}$}
\affiliation{
$^1$Department of Physics, Okayama University, Okayama 700-8530, Japan\\
$^2$Institute of Physics, Chinese Academy of Sciences, and Beijing National Laboratory for Condensed Matter Physics, Beijing 100190, China
}
\date{\today}

\begin{abstract}
	We report the $^{207}$Pb nuclear magnetic resonance (NMR) measurements on polycrystalline sample of PbTaSe$_2$
	with noncentrosymmetric crystal structure and topological electronic band.
	The nuclear spin-lattice relaxation rate $1/T_1$ shows a suppressed coherence peak below 
	the superconducting transition temperature $T_{\rm c}$ = 4.05 K and
	decreases as an exponential function of temperature.
	The penetration depth derived from the NMR spectrum is almost temperature independent below $T$ = 0.7 $T_{\rm c}$.
	The Knight shift $K$ decreases below $T_{\rm c}$.
	These results suggest spin-singlet superconductivity with a fully-opened gap $2\Delta = 3.5$ $k_{\rm B}T_{\rm c}$ in PbTaSe$_2$.
		
\end{abstract}

\maketitle
\section{introduction}
In superconductors with an inversion center in the crystal structure, 
either spin-singlet even-parity or spin-triplet odd-parity superconducting state is realized.
However, if the spatial inversion symmetry is broken, parity is no longer a conserved quantum number, 
then a spin-singlet and spin-triplet mixed superconducting state becomes possible.
In a noncentrosymmetric (NCS) system, an antisymmetric spin-orbit coupling (ASOC) interaction is induced and 
the parity-mixing extent is determined by the strength of the ASOC \cite{Gorkov_PhysRevLett.87.037004,Frigeri_PhysRevLett.92.097001}.
In some NCS superconductors, such as 
Ce(Pt,Ir)Si$_3$ \cite{Bauer_PhysRevLett.92.027003,Mukuda_PRL}, 
Li$_2$(Pd,Pt)$_3$B \cite{Li2Pd3B,Nishiyama_PhysRevB.71.220505,Li2Pt3B,Nishiyama_PhysRevLett.98.047002,Yuan_PhysRevLett.97.017006},
Mg$_{10}$Ir$_{19}$B$_{16}$ \cite{Mg10Ir19B16,Tahara_PhysRevB.80.060503},
Mo$_3$Al$_2$C  \cite{Mo3Al2C_transport_Bauer}, Re$_6$Zr \cite{ZrRe6,ZrRe6_Matano}, and so on,
 interesting properties have been reported. In all NCS compounds with weak electron correlations,
however, no striking evidence for a strong parity-mixing was observed except for
 Li$_2$(Pd$_{1-x}$Pt$_x$)$_3$B\cite{Nishiyama_PhysRevLett.98.047002,Yuan_PhysRevLett.97.017006},
where the ASOC is abruptly enhanced beyond $x$ = 0.8 due to a large tilting of octahedron-octahedron angle \cite{Harada_PhysRevB.86.220502}.

Recently, studies of NCS materials have entered a new stage for  the topological aspects
of the superconductivity\cite{sato_PhysRevB.79.094504,tanaka_PhysRevLett.105.097002}.
In topological superconductors, the existence of Majorana zero modes was suggested in the vortex core or on the edge \cite{fu_PhysRevLett.100.096407}, which has a bright application perspective.
Some roots to bulk topological superconductors with time reversal symmetry have been suggested.
One is doped topological insulators\cite{Fu_PhysRevLett.105.097001}. 
Indeed, spin-triplet odd-parity superconductivity has been observed in the doped topological insulator Cu$_x$Bi$_2$Se$_2$, 
establishing it as a topological superconductor with inversion symmetry \cite{MatanoKrienerSegawaEtAl2016}.
Also, superconducting Dirac/Wyle semimetals can be topological superconductors \cite{Haldane_PhysRevLett.120.067003,Kobayashi_PhysRevLett.115.187001}.
Another way to obtain a topological superconducting state is through NCS superconductors,
which can have a non-zero topological invariant\cite{Sato_PhysRevB.83.224511}.
In addition to Li$_2$(Pd$_{1-x}$Pt$_x$)$_3$B, materials such as LaPtBi \cite{nature-mat-9-541,PhysRevLett.105.096404} and BiPd \cite{SunEnayatMaldonadoEtAl2015} 
have recently been proposed as candidates, but clear evidence of unconventional superconductivity has not been found so far \cite{matano_JPSJ.82.084711}.

\pb\ has a layered hexagonal unit cell with the space group of $P\bar{6}m2$.
Although the compound 
has been known since 1980 \cite{EPPINGA1980121,EPPINGA1981174}, 
superconductivity was reported only recently\cite{PbTaSe2_hinetsu_PhysRevB.89.020505}. 
Angle-resolved photoemission spectroscopy measurements data showed the existence of topological nodal lines near the Fermi level \cite{BianChangSankarEtAl2016}, 
which are protected by a reflection symmetry of the space group. 
Thus, \pb\ has both aspects of NCS superconductors and topological metals.
Although $s$-wave gap was suggested by the specific heat and the penetration depth measurement \cite{PbTaSe2_Hc2,PbTaSe2_uSR_PhysRevB.93.060506},
the upper critical field $H_{{\rm c}2}$ at low-temperature shows an unconventional upward curvature as a function of
temperature \cite{PbTaSe2_Hc2_upward_PhysRevB.93.054520}.
To investigate  the superconducting state of \pb\ in more detail, further microscopic experiments are needed.

In this paper, we report $^{207}$Pb nuclear magnetic resonance (NMR) measurements of polycrystalline sample of \pb.
In the temperature dependence of spin-lattice relaxation rate \tl, we find a suppressed coherence peak just below \tc,
 and \tl\ decreases exponentially at low temperatures.
The penetration depth derived from the linewidth of the NMR spectrum shows a temperature-independent behavior below $T$=0.7 \tc.
The Knight shift decrease below \tc.
These results are consistent with a superconducting state with a full gap and spin-singlet symmetry.

\section{EXPERIMENTAL}
Polycrystalline sample of \pb\ was prepared by the solid-state reaction method using a sealed evacuated quarts tube.
Elemental starting materials of Pb (99.9999\% purity), Ta (99.9\%), and Se (99.999\%) with a stoichiometric ratio were placed in a quarts tube,
sealed under vacuum, and heated in 800$^{\rm o}$C for one week.
The melted sample was crashed into powder for X-ray diffraction (XRD) and NMR measurement.
XRD measurement are performed with Cu $K_\alpha$ radiation at room temperature. 
The $T_{\rm c}$ was determined by measuring the inductance of the in situ NMR coil. 
NMR measurements were carried out by using a phase-coherent spectrometer. 
The $^{207}$Pb-NMR spectra were obtained by the fast Fourier
transform of the spin-echo obtained with a standard $\pi/2-\tau - \pi$ sequence. 
The Knight shift $K$ was calculated using nuclear gyrometric ratio $\gamma_{\rm N}$ = 8.874 MHz/T for $^{207}$Pb.
In order to minimize the reduction of $T_{\rm c}$ by the applied magnetic field, 
measurements were performed at a small field of $H_0 $= 0.19 T.
The $T_1$ was measured by using a single saturating pulse, 
and determined by fitting the recovery curve of the nuclear magnetization to a single exponential function:
$\frac{M_0-M(t)}{M_0} = \exp(-\frac{t}{T_1})$, 
where $M_0$ is the nuclear magnetization in the thermal equilibrium and $M(t)$ is the nuclear magnetization at a time $t$ after the saturating pulse.
The recovery curve was well fitted with a unique $T_1$ component for all temperature range.
\begin{figure}[htbp]
	\includegraphics[clip,width=80mm]{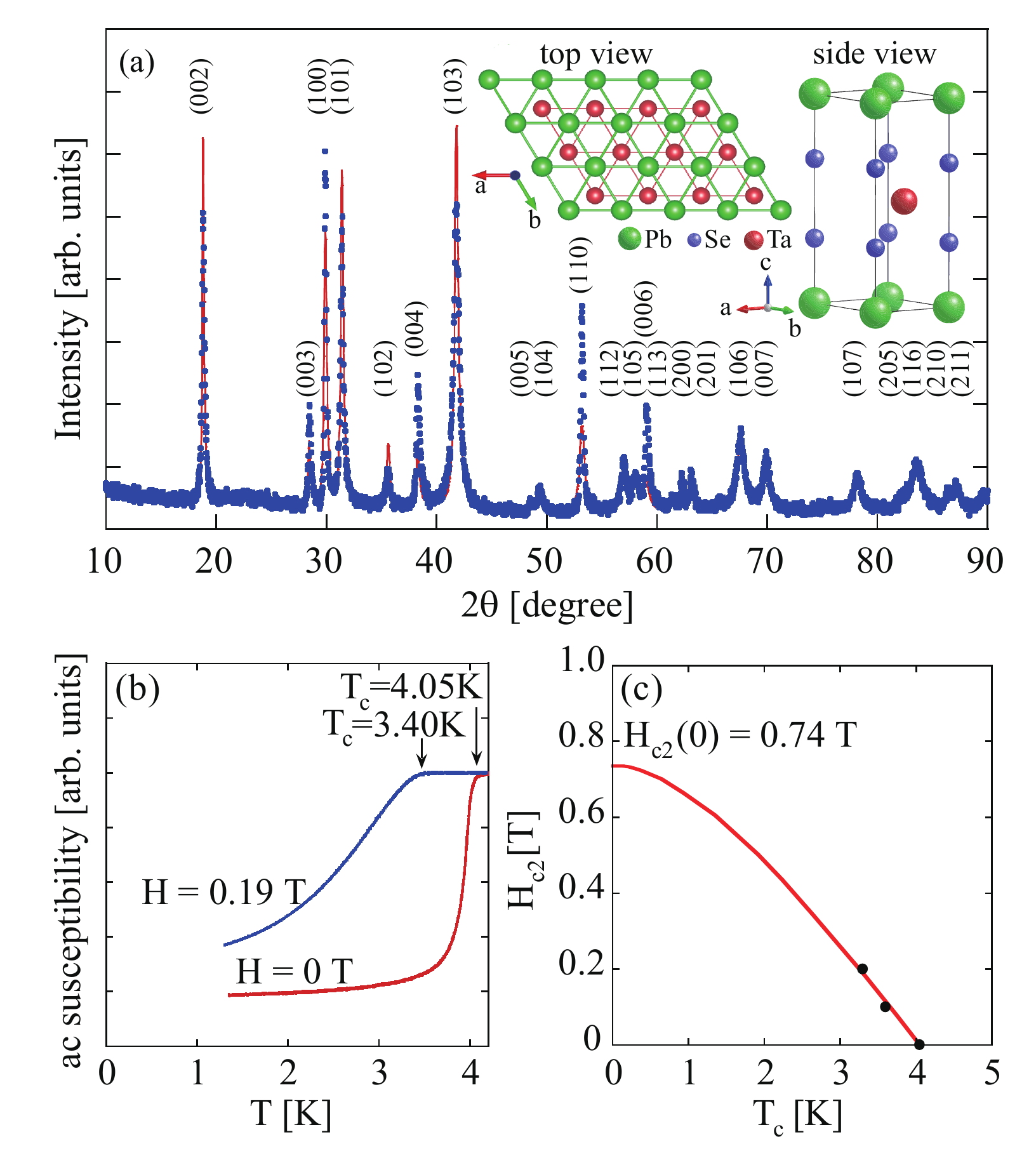}
	\caption{\label{x-ray}
		(Color online) (a) The powder XRD pattern for \pb. 
		The dotted line is the observed data and red line shows the theoretical fitting obtained by the Rietveld method.
		Reflections are indexed with respect to the space group of $P\bar{6}m2$.
		The inset shows the crystal structure of \pb. 
		(b) Ac susceptibility measured using the in situ NMR coil in $H$ = 0 and 0.19 T. 
		The arrow indicates \tc\ for each field.
		(c) Temperature dependence of the upper critical field $H_{{\rm c}2}(T)$,
		the solid curve is the fitting to the WHH theory.
	}	
\end{figure}

\section{results and discussions}
Figure 1(a) shows the XRD pattern for \pb.
The result can be fitted by the Rietveld method, with no peaks from secondary phase.
We obtained the lattice constant $a$ = 3.436 \AA\ and $c$ = 9.375 \AA, 
which are close to the previous report of $a$ = 3.44 \AA\ and $c$ = 9.35 \AA  \cite{EPPINGA1980121}.
Figure 1(b) shows the ac susceptibility verses temperature for the zero field and an applied magnetic field $H$ = 0.19 T, respectively.
The \tc\ for zero magnetic field is 4.05 K, which is higher than 3.7 - 3.9 K reported previously \cite{PbTaSe2_hinetsu_PhysRevB.89.020505,PbTaSe2_uSR_PhysRevB.93.060506,PbTaSe2_Hc2_upward_PhysRevB.93.054520}.
The value of \tc\ is reduced to 3.40 K at $H$ = 0.19 T. 
The temperature dependence of the upper critical field $H_{c2}(T)$ is plotted in Fig. 1(c)
, with the solid curve for Werthamer-Helfand-Hohenberg (WHH) theory \cite{WHH}.
\begin{figure}[htbp]
	\includegraphics[clip,width=80mm]{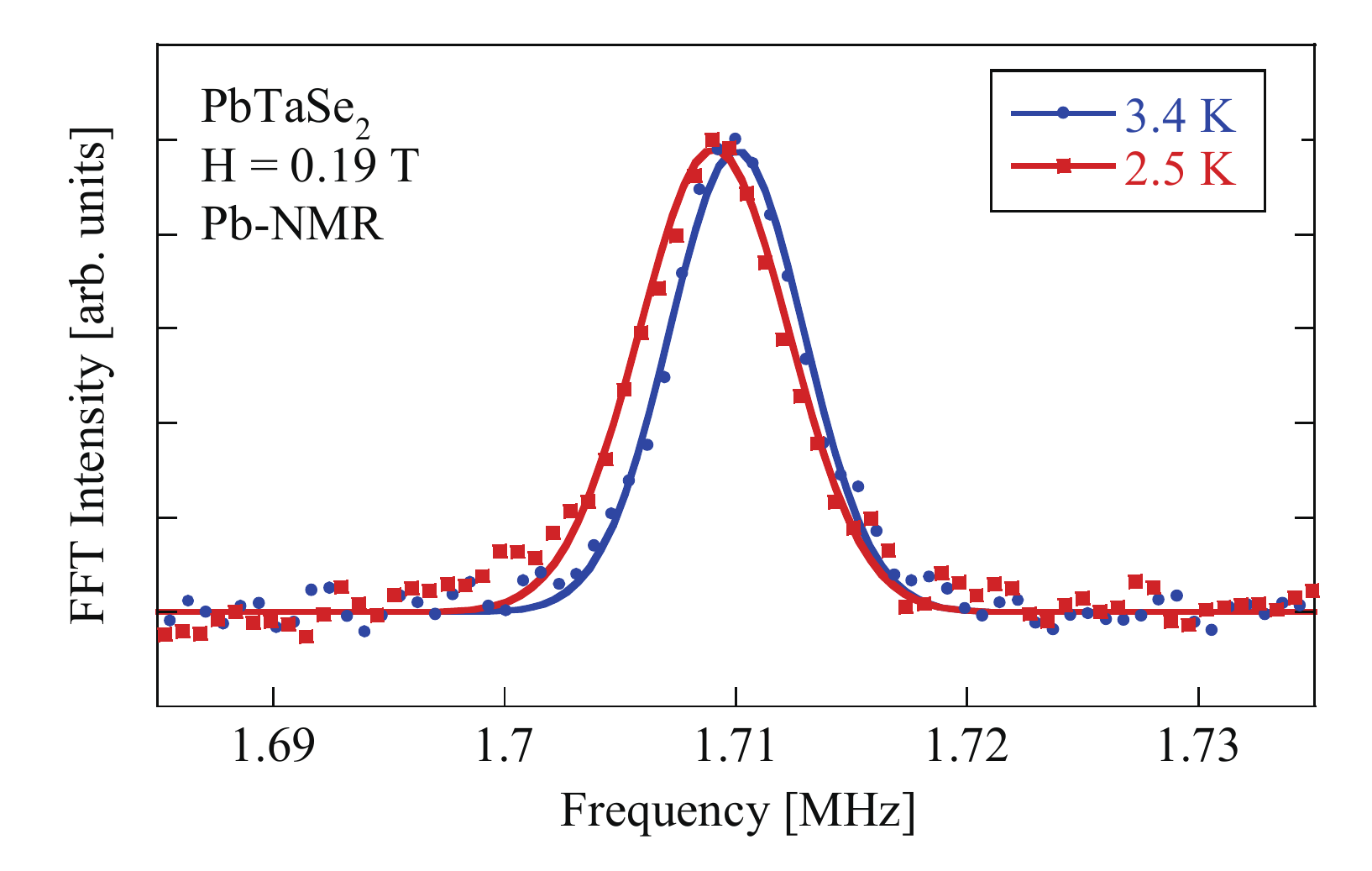}
	\caption{\label{spec}(color online) The $^{207}$Pb-NMR spectra measured in a magnetic field of $H$ = 0.19 T above and below $T_c(H)$ = 3.4 K respectively.
	The solid curves are Gaussian fits to each spectrum.	
	}
\end{figure}

Figure \ref{spec} shows the $^{207}$Pb ($I = 1/2$) NMR spectrum above and below \tc\, respectively.
The spectra can be fitted by a single Gaussian.
\begin{figure}[htbp]
	\includegraphics[clip,width=80mm]{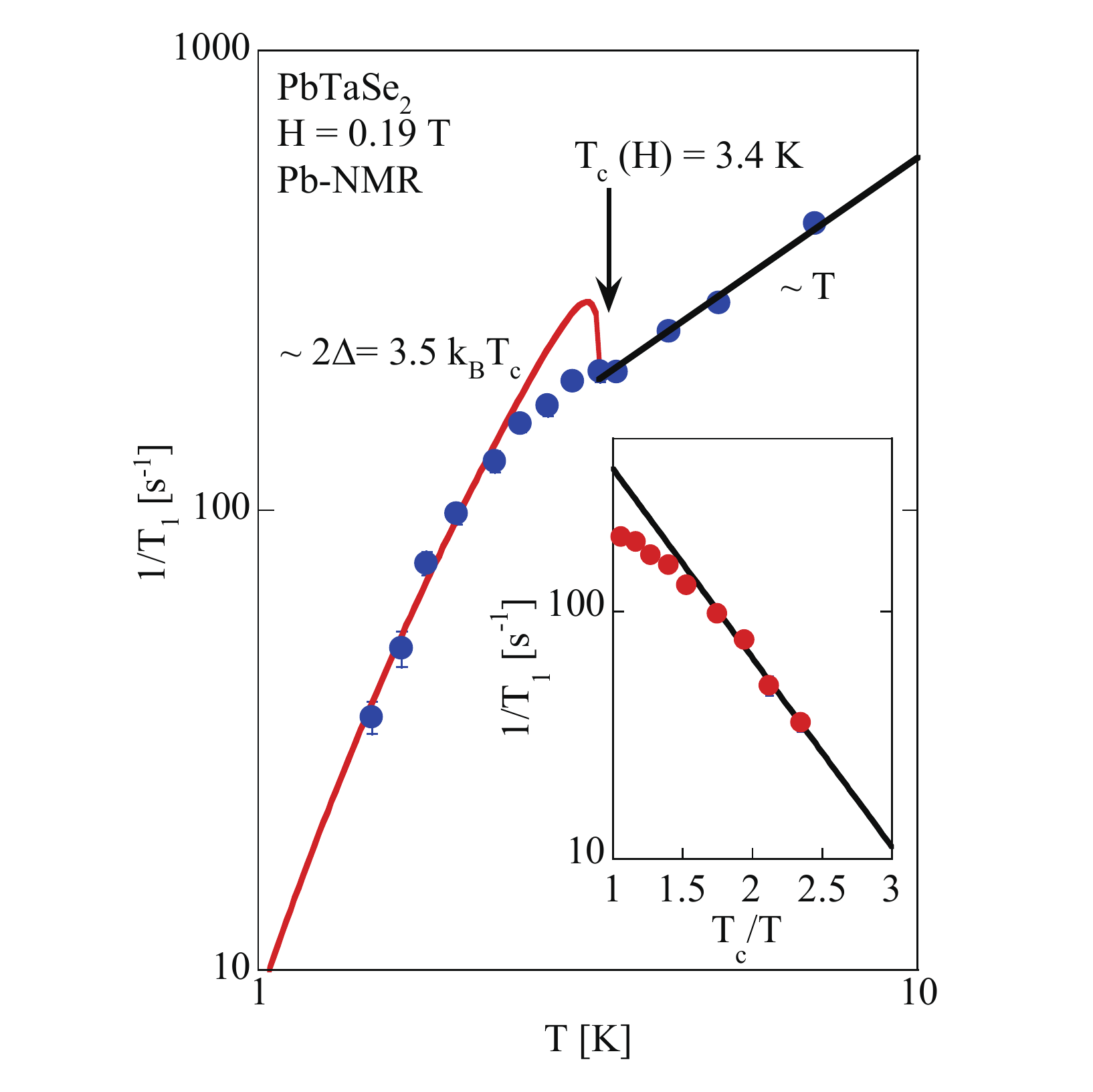}
	\caption{\label{T1}
		(color online) Temperature dependence of the spin-lattice relaxation rate $1/T_1$ measured by $^{207}$Pb-NMR.
		The straight line above \tc\ represents the $T_1T$ = const relation.
		The solid curve below \tc\ is a calculation assuming the $s$-wave gap function (see text).
		The inset shows the semilogarithmic plot of \tl\ versus $T_{\rm c}/T$.
		The solid line in the inset represents the relation $1/T_1 \propto e^{\left(-\frac{\Delta_0}{k_{B}T}\right)}$
	}
\end{figure}
Figure \ref{T1} shows the temperature dependence of \tl\ measured at the peak of the NMR spectrum at $H$ = 0.19 T.
As seen in the figure, \tl\ varies in proportion to the temperature ($T$) above \tc, as expected for conventional metals, indicating no electron-electron interaction.
Below \tc, \tl\ shows a suppressed coherence peak (Hebel-Slichter peak), but decreases exponentially at low temperatures.
The inset to  Fig. \ref{T1} shows a semilogarithmic plot of \tl\ as a function of $T_{\rm c}/T$, indicating more visually that \tl\ decays exponentially with respect to temperature.
The spin-lattice relaxation rate in the superconducting state $1/T_{1S}$ is expressed as
\begin{eqnarray}\label{t1cal}
\frac{T_{1N}}{T_{1S}}=\frac{2}{k_BT} 
\iint\left(1+\frac{\Delta^2}{EE'}\right)N_S(E)N_S(E')  \nonumber \\
\times f(E)\left[1-f(E')\right]\delta(E-E')dEdE',
\end{eqnarray}
where $1/T_{1N}$ is the relaxation rate in the normal state, $N_S(E)$ is the superconducting density of
states (DOS), and $f(E)$ is the Fermi distribution function.
$C =1+ \frac{\Delta^2}{EE'}$ is the coherence factor with $\Delta$ being the superconducting gap.
To perform the calculation of eq. \ref{t1cal}, we follow Hebel to convolute $N_S(E)$ with a broadening function $B(E)$ \cite{Hebel},
which is approximated with a rectangular function centered at $E$ with a height of 1/2$\delta$.
The solid curve below \tc\ shown in Fig. \ref{T1}  is a calculation based on the BSC gap function with $2\Delta = 3.5 $ \kb, $r\equiv\Delta(0)/\delta=1.8$.
It fits the experimental data except for just below \tc. The parameter $2\Delta$ is close to the BCS value of 3.53 \kb.

We comment on the lack of coherence peak. 
First of all, non $s$-wave superconducting state produces no coherence peak.
Even in an $s$-wave superconductor,
the height of the coherence peak can be suppressed by many factors such as a large applied magnetic field,  
anisotropy of superconducting gap\cite{MacLaughlin19761}, multiple superconducting gaps\cite{Matano_BaFeAs},
large nuclear electric quadrupole moment \cite{LiZheng}, and phonon damping in the strong coupling regime\cite{kitaoka1992272,inpress}.
Since $2\Delta =$ 3.5 \kb\ indicates a weak-coupling superconductor, phonon damping is unlikely as a origin.
$^{207}$Pb ($I=1/2$) has no electric quadrupole moment, so the effect of the electric quadrupole interaction is also excluded.
At the moment, an anisotropic superconducting gap or the multi-band electronic structure is likely to be responsible for the suppressed coherence peak 
in the case of $s$-wave symmetry.
Indeed, the multi-gap superconductivity is suggested from the thermal conductivity and the upper critical field measurements in Refs. 
\onlinecite{PhysRevB.93.020503,PbTaSe2_Hc2_upward_PhysRevB.93.054520}.
In any case, the exponential decay of \tl\ suggests a fully opened gap.

\begin{figure}[htbp]
	\includegraphics[clip,width=80mm]{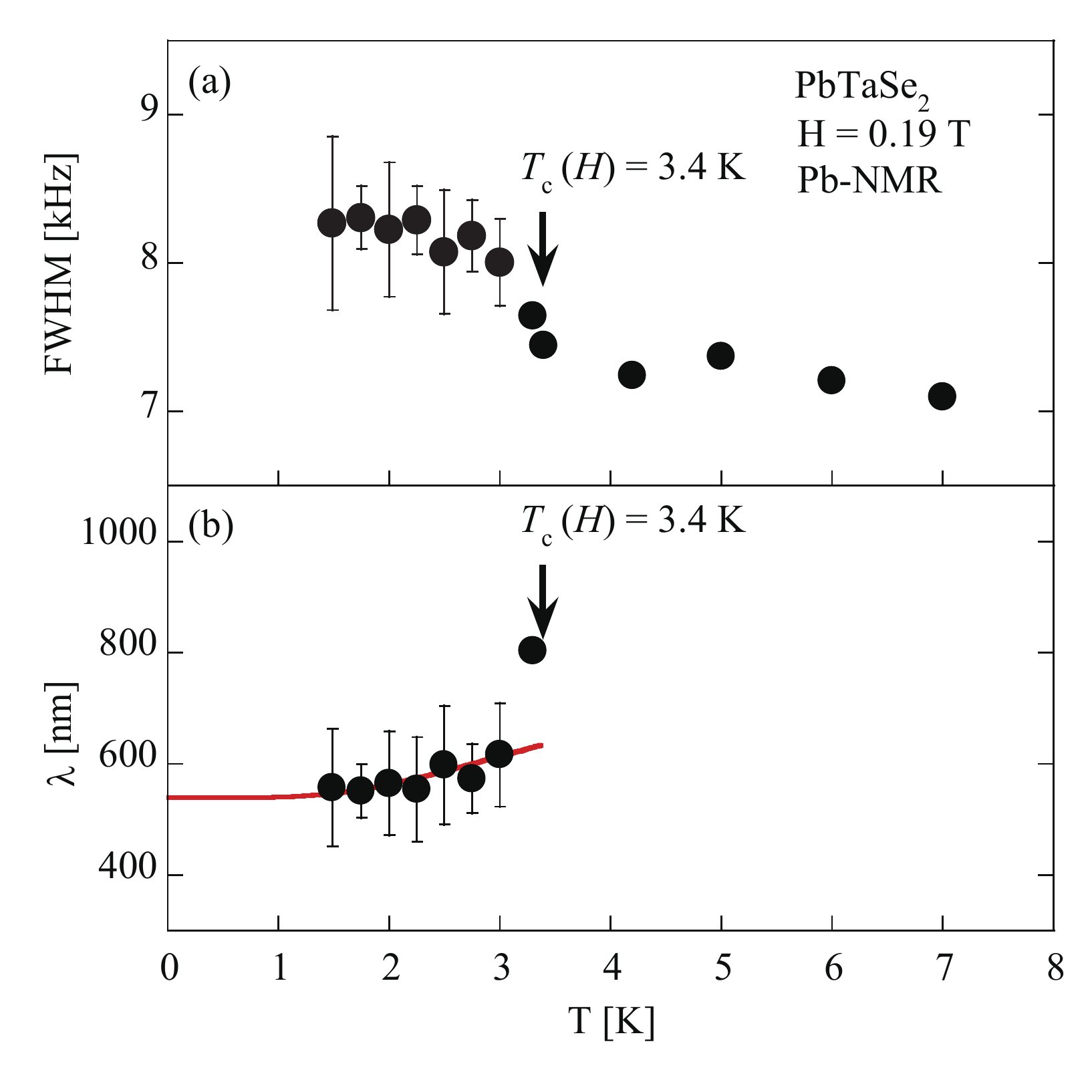}
	\caption{\label{fwhm}(color online) (a) The temperature variation of  the full width at the half maximum (FWHM) for the $^{207}$Pb-NMR spectrum.
		(b) The temperature dependence of penetration depth $\lambda$ calculated from the FWHM (see text for detail).
	}
\end{figure}
\begin{figure}[htbp]
	\includegraphics[clip,width=80mm]{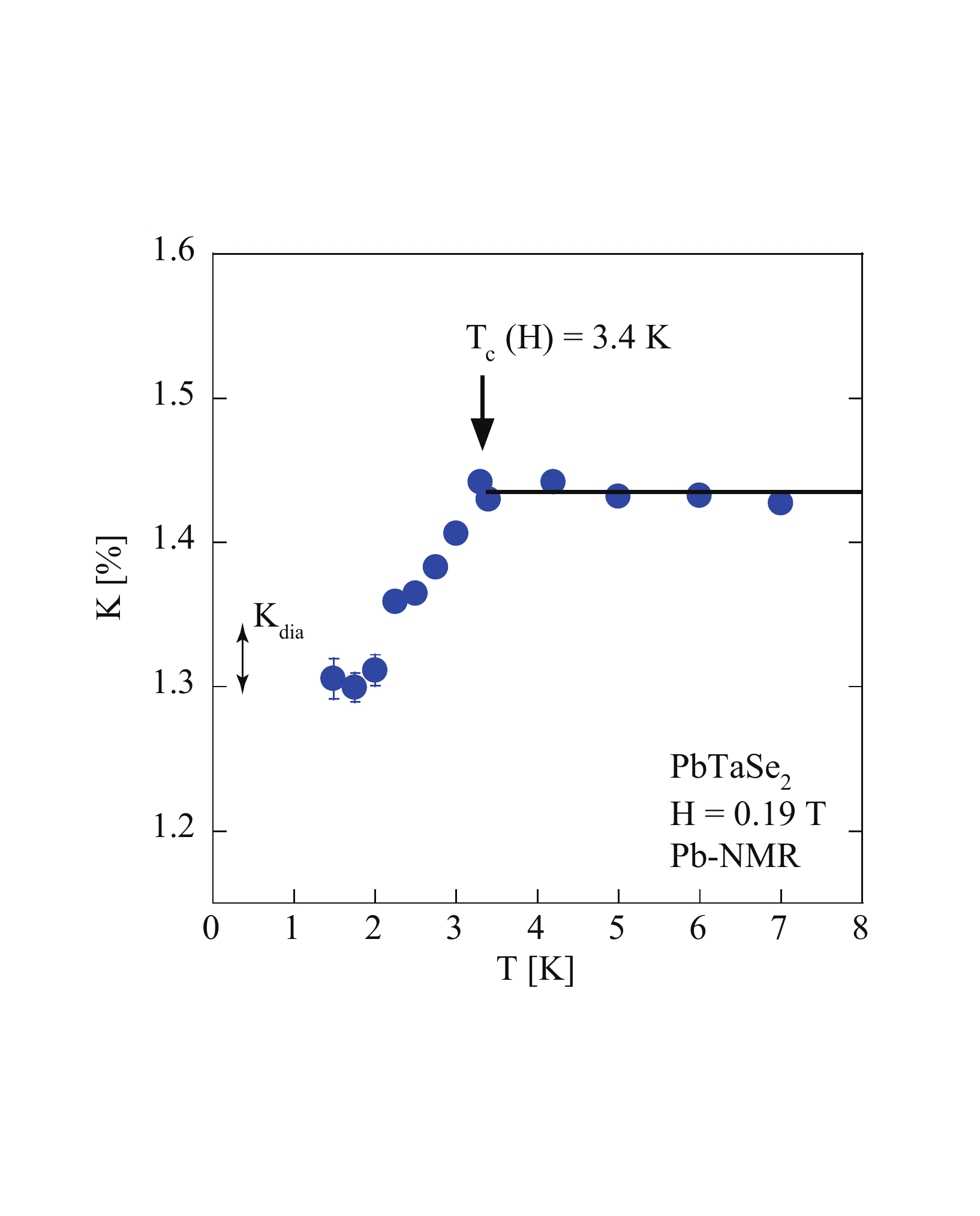}
	\caption{\label{k}
		(color online) The Knight shift versus temperature for \pb\ measured at $H = 0.19$ T. The size of diamagnetic contribution  $K_{\rm dia}$ in the zero-temperature limit due to the formation of vortex lattice is shown in the figure.
	}
\end{figure} 
The temperature dependence of the penetration depth derived from the NMR spectra supports the conclusion.
The full width at half maximum (FWHM) as a function of temperature is plotted in Fig. \ref{fwhm}(a).
Above \tc, the FWHM is temperature independent. 
However, the FWHM increases below $T{_c}(H)$, due to a distribution of the magnetic field in a superconductor in the vortex state\cite{Voltex_PhysRevLett.88.077003}.
The difference of the FWHM  between temperatures below and above \tc\
is related to the London penetration depth $\lambda$ as the following\cite{Voltex_PhysRevB.37.2349}, 
\begin{eqnarray}
{\textstyle  \sqrt{FW\!H\!M(T=0)^2-FW\!H\!M(T\geq T_c)^2 }}  = 0.0609\gamma_n\frac{\phi_0}{\lambda^2}.\nonumber
\\
\textstyle 
\end{eqnarray}
Here, $\gamma_n$ is the gyromagnetic ratio for a nucleus,  
$\phi_0 = 2h/e = 2.07 \times 10^{-7}$Oe/cm$^2$ is the quantized magnetic flux.
Our result shows that the penetration depth is $T$-independent below $T$=0.7 \tc, indicating a full gap superconducting state.
If there are nodes in the superconducting gap function, $\lambda(T)$ will  show a $T$-linear temperature dependence at low temperatures.
For a fully gapped superconductor, the temperature dependence of $\lambda$ for $T/T_c<0.5$ is described as following \cite{lambda},
\begin{eqnarray}
\lambda(T) = \lambda(0)\left[1+\sqrt[]{\frac{\pi\Delta}{2k_BT}}\exp\left(-\frac{\Delta}{k_BT}\right) \right].
\end{eqnarray}
The solid line below \tc\ in Fig. \ref{fwhm}(b) is the calculation by assuming a fully gapped superconducting state with $\Delta$($T=0$) obtained from the $T_1$ result 
and $\lambda(0)$ is calculated to be 550 nm.
This value is a little larger than previous $\mu$SR report in the applied magnetic field of $\lambda$(0) = 180 nm in 0.040 T and 140 nm in 0.025 T \cite{PbTaSe2_uSR_PhysRevB.95.224506}. 
The temperature below which $\lambda(T)$ becomes $T$-independent is also slightly higher compared to Ref. \onlinecite{PbTaSe2_uSR_PhysRevB.95.224506}.
Differences in measuring conditions may be one of the causes.
In $\mu$SR study, the magnetic field was applied parallel to the $c$-axis for which
$H_{{\rm c}2}^{\rm c}$ was calculated to be 0.1 T\cite{PbTaSe2_uSR_PhysRevB.95.224506},
which is much lower than the applied magnetic field in this study. 
An anisotropy of physical properties could be another cause.  
When the Knight shift is anisotropic, the observed spectrum in the polycrystalline sample will be broadened.
In calculating $\lambda$ from the FWHM, such broadening was not taken into account.
At any rate, the temperature dependence of $\lambda$ 
together with the result of \tl\ are consistent with recent the specific heat and the London penetration depth measurement \cite{PbTaSe2_Hc2,PbTaSe2_uSR_PhysRevB.93.060506}.

Figure \ref{k} shows the Knight shift, $K$, as a function of temperature.
Above \tc, the shift is $T$ independent, while it decreases below \tc.
Generally, the Knight shift in the superconducting state is expressed as,
\begin{eqnarray}
K &&= K_{\rm orb}+K_{\rm s} + K_{\rm dia},\\
K_s &&= A_{\rm hf}\chi_{\rm s},\\
\chi_{\rm s} &&= -4\mu^2_{\rm B}\int N_{\rm S}(E)\frac{\partial f(E)}{\partial E}dE,
\end{eqnarray}
where $K_{\rm orb}$ is the contribution due to orbital susceptibility which is $T$-independent, 
$A_{\rm hf}$ is the hyperfine coupling constant, $\chi_{\rm s}$ is the spin susceptibility,
and $K_{\rm dia}$ is contribution from diamagnetism in the vortex state.
The $K_{\rm dia}$ is calculated by using following equation for diamagnetic field $H_{\rm dia}$ \cite{gennes},
\begin{eqnarray}
H_{\rm dia} = H_{\rm c1} \frac{\ln \left(\frac{\beta d}{\sqrt{e}\xi}\right)}{\ln \frac{\lambda}{\xi}}
\end{eqnarray}
here, $\beta$ is 0.38 for triangular lattice of vortex, $\xi$ is the coherence length, $\lambda$ is the London penetration depth.
For $\xi$ and $\lambda$, the values ​​obtained from the measurement of $H_{\rm c2}$ and the FWHM were used.
As a result, $K_{\rm dia}$ was calculated to be -0.05 \% at most.
Even if $K_{\rm dia}$ is taken into consideration, the Knight shift decreases substantially below \tc,
which is consistent with the spin-singlet Cooper pairing.
The large finite value of $K$ at the lowest temperature is probably due to
 $K_{\rm orb}$. Further work is needed to determine  $K_{\rm orb}$ in the future.
 In addition,
in the presence of strong spin-orbit coupling, 
spin-orbit scattering can also cause a large spin susceptibility in the superconducting state\cite{PhysRevLett.3.262}. 
This is very likely in the present case  as Pb is a heavy elements and the sample is powdered\cite{PhysRevLett.3.262}.

\section{summary} 
In summary, a polycrystalline sample of noncentrosymmetric superconductor \pb\
was investigated by using $^{207}$Pb-NMR method.
We find that the spin-lattice relaxation rate \tl\ exhibits an exponential behavior with respect to temperature, with a suppressed coherence peak just below \tc, which suggests a fully opened gap. 
The temperature variation of the penetration depth derived from the NMR spectrum supports this conclusion. 
The spin susceptibility measured by the Knight shift decreases below \tc, 
suggesting that the Cooper pairs are in a singlet state. 

\begin{acknowledgments}
We thank G. F. Chen for collaboration using a single  crystal in the early stage of this work.
This work was supported by the JSPS  Grants (Nos. JP15H05852, JP16H04016, and JP17K14340).
\end{acknowledgments}

\end{document}